\documentclass[12pt]{article}

\begin{document}

\author{T\^{a}nia Tom\'{e} and M\'{a}rio J. de Oliveira \\
%EndAName
Instituto de F\'{\i}sica\\
Universidade de S\~{a}o Paulo\\
Caixa Postal 66318\\
05315-970 S\~{a}o Paulo, S\~{a}o Paulo, Brazil}
\title{Renormalization group of the Domany-Kinzel cellular automaton}
\maketitle

\begin{abstract}
We apply the dinamically driven renormalization group to study the critical
properties of the Domany-Kinzel probabilistic cellular automaton. To
preserve the absorbing state clusters with at least one site occupied are
renormalized into a one occupied site and clusters with all sites empty are
renormalized into a one empty site. We have obtained the phase diagram as
well as the critical exponent related to the spatial correlation length.
\end{abstract}

\newpage\ 

\section{Introduction}

Among the variety of irreversible models that describe a nonequilibrium
phase transition from an active state to an absorbing state, the
Domany-Kinzel probabilistic cellular automaton \cite{dom} together with the
contact process \cite{ligg} are perhaps the simplest ones. According to
Janssen \cite{jan} and Grassberger \cite{grass}, any one component model
displaying a transition from an absorbing to an active state belongs in the
same universality class as direct percolation. In fact, the Domany-Kinzel
cellular automaton is equivalent to direct percolation.

The Domany-Kinzel probabilistic cellular automaton is defined as follows. In
a one dimensional lattice each site can be in one of two states, either
empty or occupied. At each time step the sites are updated simultaneously
and independently. The state of a given site depends only on the states of
two neighboring sites in the previous time step. The rules are: 1) if the
two sites are both empty, the given site will be empty, 2) if one site is
occupied and the other is empty, the given site will be occupied with
probability $p_1,$ and 3) if the two sites are both occupied, the given site
will be occupied with probability $p_2.$ Due to the rule number 1, the state
with all sites empty will be an absorbing state.

In the stationary state the Domany-Kinzel automaton exhibits two states: the
absorbing state with all sites empty and an active state with a nonzero
density of occupied sites. For small values of $p_1$ the system is in the
absorbing state. When the parameter $p_1$ is increased the absorbing state
becomes unstable, for sufficiently large values of $p_1,$ given rise to an
active state. The phase transition from the absorbing to the active state is
a continuous transition and belongs to the same universality class as direct
percolation.

Renormalization group (RG) techniques in real space have been used
successfully in studying the critical behavior of equilibrium statistical
models since the work of Niemeyer and van Leeuwen \cite{niem}. In these
techniques small clusters of sites are renormalized into just one site
according to certain rules. For the case of the Ising model with the up-down
symmetry, the most important rule is the majority rule which preserves the
up-down symmetry of the model at each step of the renormalization. Any RG
scheme should be set up in a way to preserve the symmetries and the main
properties of the model studied. In the present case the most important
property is the existence of an absorbing state which should then be
preserved by the RG trasformation. To this end, we have constructed a real
space RG such that a cluster with at least one site occupied is renormalized
into a one occupied site. Only clusters with all sites empty are
renormalized into a one empty site.

For systems in equilibrium, that is, for systems defined by a Hamiltonian
the appropriate RG space is the space of the parameters, or coupling
constants, that define the Hamiltonian. A sequence of RG transformations
correspond to a trajectory in this space. Systems far from equilibrium, that
is, systems that lack detailed balance, on the other hand, are not defined
by Hamiltonians but by the dynamic rules or equivalently by the transition
probabilities that govern its time evolution. In this case the appropriate
RG space is the space of parameters that define the transition probabilities.

Here we use a real space RG \cite{maz1} \cite{maz2} \cite{haake} that
renormalizes the transition probabilities so that one obtains a RG
transformation in the space spanned by the parameters that define the
transition probabilities. In the case of the Domany-Kinzel cellular
automaton studied here this space will be the space defined by the
parameters $p_1,$ and $p_2.$ The scheme we use is an application of the
dynamically driven renormalization group introduced by Pietronero et al. 
\cite{pietro} and Vespignani et al. \cite{vespi1} \cite{vespi2} to study
systems in critical states. A study of the Domany-Kinzel by means of a
phenomenological RG has been done by Bagnoli et al. \cite{bagno}.

The RG transformation was accomplished by using cells of size $b=2$ and by
using a series of approximations to the stationary state. As we increase the
order of the approximation better values were obtained for the universal
critical exponent as well as for the nonuniversal critical quantities. Our
best value for the critical exponent $\nu _{\perp }$ related to spatial
correlation length is $\nu _{\perp }=1.013$ which should be compared to $\nu
_{\perp }=1.100\pm 0.005$ obtained from transfer matrix methods \cite{dom}.
The critical value of $p_1$ along the line $p_2=0$ is $p_c=0.792$ which is
close to $p_c=0.799\pm 0.002$ obtained from simulations \cite{zebende}.\ 

\section{RG transformation}

Consider a one dimensional lattice of $N$ sites in which each site can be in
one of two states, either empty ($\sigma _i=0$) or occupied ($\sigma _i=1$).
At each time step the sites are updated simultaneously, and independently,
so that the transition probability $W(\sigma |\sigma ^{\prime })$ from state 
$\sigma ^{\prime }=(\sigma _1^{\prime },\sigma _2^{\prime },...,\sigma
_N^{\prime })$ to state $\sigma =(\sigma _1,\sigma _2,...,\sigma _N)$ is
given by 
\begin{equation}
W(\sigma |\sigma ^{\prime })=\prod_iw(\sigma _i|\sigma _i^{\prime },\sigma
_{i+1}^{\prime }),  \label{1}
\end{equation}
where 
\begin{equation}
w(1|00)=0,\qquad w(1|10)=w(1|01)=p_1,\qquad w(1|11)=p_2.  \label{2}
\end{equation}
These rules completely defines the Domany-Kinzel probabilistic cellular
automaton since 
\begin{equation}
w(0|\sigma _i^{\prime },\sigma _{i+1}^{\prime })+w(1|\sigma _i^{\prime
},\sigma _{i+1}^{\prime })=1.  \label{3}
\end{equation}

We remark that the expression (\ref{1}) implies that, given the state $%
\sigma ^{\prime },$ the variables $\sigma _i$ are independent random
variables which in turn is equivalent to say that each site should be
updated independently of the others, as desired. We call this property of
the cellular automaton \emph{one-site independence}.

In the stationary state the following condition 
\begin{equation}
P(\sigma )=\sum_{\sigma ^{\prime }}W^\ell (\sigma |\sigma ^{\prime
})P(\sigma ^{\prime })  \label{12}
\end{equation}
holds for any value of $\ell ,$ where $P(\sigma )$ is the stationary
probability. If we use the notation $P_\ell (\sigma ,\sigma ^{\prime })$ for
the probability of having state $\sigma ^{\prime }$ at a given time and
state $\sigma $ at a $\ell $ time steps later then 
\begin{equation}
P_\ell (\sigma ,\sigma ^{\prime })=W^\ell (\sigma |\sigma ^{\prime
})P(\sigma ^{\prime }).  \label{13}
\end{equation}
Similarly, 
\begin{equation}
\widetilde{P}(\tau ,\tau ^{\prime })=\widetilde{W}(\tau |\tau ^{\prime })%
\widetilde{P}(\tau ^{\prime })  \label{14}
\end{equation}
is the analog equation for the renormalized system where $\widetilde{W}(\tau
|\tau ^{\prime })$ is the (one-step) renormalized transition probability we
are looking for and $\tau =(\tau _1,\tau _2,$ $...,\tau _{N^{\prime }}),$
with $N^{\prime }=N/b,$ denotes a renormalized state.

Let $\mathcal{R}(\tau |\sigma )$ be a conditional probability of the
renormalized state $\tau $ given the state $\sigma ,$ which has the
following properties 
\begin{equation}
\mathcal{R}(\tau |\sigma )\geq 0,\qquad \qquad \sum_\tau \mathcal{R}(\tau
|\sigma )=1.  \label{11}
\end{equation}
The RG transformations are obtained by demanding that \cite{maz1} 
\begin{equation}
\widetilde{P}(\tau ,\tau ^{\prime })=\sum_\sigma \sum_{\sigma ^{\prime }}%
\mathcal{R}(\tau |\sigma )\mathcal{R}(\tau ^{\prime }|\sigma ^{\prime
})P_\ell (\sigma ,\sigma ^{\prime })  \label{15}
\end{equation}
from which follows 
\begin{equation}
\widetilde{P}(\tau ^{\prime })=\sum_{\sigma ^{\prime }}\mathcal{R}(\tau
^{\prime }|\sigma ^{\prime })P(\sigma ^{\prime }).  \label{16}
\end{equation}
This last expression is the one used in real space RG for equilibrium
systems. Taking the ratio of the last two equations we obtain \cite{vespi2} 
\begin{equation}
\widetilde{W}(\tau |\tau ^{\prime })=\frac{\sum_\sigma \sum_{\sigma ^{\prime
}}\mathcal{R}(\tau |\sigma )W^\ell (\sigma |\sigma ^{\prime })\mathcal{R}%
(\tau ^{\prime }|\sigma ^{\prime })P(\sigma ^{\prime })}{\sum_{\sigma
^{\prime }}\mathcal{R}(\tau ^{\prime }|\sigma ^{\prime })P(\sigma ^{\prime })%
},  \label{17}
\end{equation}
which is the desired expression for the renormalized transition probability.
Due to the presence of the stationary probability on the right hand side of (%
\ref{17}), this is not a transformation that involves only the transition
probability. However, if we use an approximation for the steady state $%
P(\sigma ),$ such as a mean field approximation, then equation (\ref{17})
provides a well defined RG transformation $W\rightarrow \widetilde{W}
$ from the old $W$ to the new $\widetilde{W}$ transition probability.

It is interesting to write equation (\ref{17}) in the form 
\begin{equation}
\widetilde{W}(\tau |\tau ^{\prime })=\sum_\sigma \sum_{\sigma ^{\prime }}%
\mathcal{R}(\tau |\sigma )W^\ell (\sigma |\sigma ^{\prime })\mathcal{S}%
(\sigma ^{\prime }|\tau ^{\prime })  \label{17a}
\end{equation}
where 
\begin{equation}
\mathcal{S}(\sigma ^{\prime }|\tau ^{\prime })=\frac{\mathcal{R}(\tau
^{\prime }|\sigma ^{\prime })P(\sigma ^{\prime })}{\sum_{\sigma ^{\prime }}%
\mathcal{R}(\tau ^{\prime }|\sigma ^{\prime })P(\sigma ^{\prime })}.
\label{17b}
\end{equation}
From this last expression it is clear that $\mathcal{S}(\sigma ^{\prime
}|\tau ^{\prime })$ is normalized with respect to the variables $\sigma
^{\prime }$ and that it can be interpreted as a conditional probability of $%
\sigma ^{\prime }$ given $\tau ^{\prime }$. In this way it becomes quite
obvious that the renormalized transition probability $\widetilde{W}(\tau
|\tau ^{\prime })$ is properly normalized with respect to the variables $%
\tau $ as it should. One important property of $\mathcal{S}$ follows from a
specific property of $\mathcal{R}$. We will see in the next section that $%
\mathcal{R}(\tau |\sigma )$ is either zero or one from which follows that $%
\mathcal{R}(\tau ^{\prime }|\sigma )\mathcal{R}(\tau |\sigma )=\delta (\tau
^{\prime },\tau )\mathcal{R}(\tau |\sigma ).$ Using this last equality we
can check that 
\begin{equation}
\sum_\sigma \mathcal{R}(\tau |\sigma )\mathcal{S}(\sigma |\tau ^{\prime
})=\delta (\tau ^{\prime },\tau ),  \label{17c}
\end{equation}
independently of $P(\sigma ).$

We remark that we should not expect, in general, that, given the state $\tau
^{\prime },$ the variables $\tau _i$ be independent. In other words, the
stochastic process defined by $\widetilde{W}(\tau |\tau ^{\prime }),$ given
by equation (\ref{17}), will not have the desired one-site independence.
This property, however, can be accomplished in an approximate way as
follows. Let us define the one-site marginal transition probability $%
\widetilde{w}(\tau _i|\tau ^{\prime })$ by 
\begin{equation}
\widetilde{w}(\tau _i|\tau ^{\prime })=\sum_\tau {}^{\prime }\widetilde{W}%
(\tau |\tau ^{\prime }),  \label{18}
\end{equation}
where the sum is over all $\tau _j$ except $\tau _i.$ Next we construct a
transition probability $\widetilde{W}_R(\tau |\tau ^{\prime })$ as being the
product of these one-site transition probabilities, that is, we define 
\begin{equation}
\widetilde{W}_R(\tau |\tau ^{\prime })=\prod_i\widetilde{w}(\tau _i|\tau
^{\prime }).  \label{19}
\end{equation}
Of course $\widetilde{W}_R(\tau |\tau ^{\prime })$ does not coincide, in
general, with $\widetilde{W}(\tau |\tau ^{\prime })$.

The final step in the renormalization procedure is to say that the
renormalized cellular automaton will be defined by the transition
probability $\widetilde{W}_R(\tau |\tau ^{\prime })$ which in turn has the
form of equation (\ref{1}). The last step is an approximation which we call
the one-site independence approximation. In this manner we obtain a
renormalized probabilistic cellular automaton, with the property of one-site
independence, whose one-site transition probability $\widetilde{w}(\tau
_i|\tau ^{\prime })$ is obtained from the original one $w(\sigma _i|\sigma
^{\prime })$ through equations (\ref{18}), (\ref{17}), and (\ref{1}). Thus
we understand the present renormalization as the transformation $w(\sigma
_i|\sigma ^{\prime })$ $\rightarrow $ $\widetilde{w}(\tau _i|\tau ^{\prime
}) $.

\section{Renormalization algorithm}

We use a renormalization $\mathcal{R}(\tau |\sigma )$ in the form of a
product which for the case $b=2$ reads 
\begin{equation}
\mathcal{R}(\tau |\sigma )=\prod_{k=1}^{N/2}R(\tau _k|\sigma _{2k},\sigma
_{2k+1}),  \label{20}
\end{equation}
where 
\begin{equation}
R(1|00)=0,\qquad R(1|10)=R(1|01)=R(1|11)=1,  \label{21}
\end{equation}
and 
\begin{equation}
R(0|00)=1,\qquad R(0|10)=R(0|01)=R(0|11)=0.  \label{22}
\end{equation}
The rule $R(0|00)=1$ ensures that the state with all sites empty will be
again an absorbing state after the renormalization transformation.

To calculate the renormalized transition probability $\widetilde{w}(\tau
_1|\tau _2,\tau _3)$ we consider a cluster of nine sites of the original
space-time lattice: two sites (labeled $1$ and $2$) at a time $\ell $, three
(labeled $3,$ $4$ and $5$) at time $\ell -1,$ and four (labeled $6,$ $7,$ $8$
and $9$) at time $\ell -2.$ The renormalization of the sites are as follows: 
$(\sigma _1,\sigma _2)\rightarrow \tau _1,$ $(\sigma _6,\sigma
_7)\rightarrow \tau _2,$ and $(\sigma _8,\sigma _9)\rightarrow \tau _3.$
Using equations (\ref{18}), (\ref{17b}), and (\ref{17a}), the renormalized
transition probability $\widetilde{w}(\tau _1|\tau _2,\tau _3)$ are
calculated by 
\[
\widetilde{w}(\tau _1|\tau _2,\tau _3)=\sum_{\sigma _1}\sum_{\sigma
_2}\sum_{\sigma _6}\sum_{\sigma _7}\sum_{\sigma _8}\sum_{\sigma _9}R(\tau
_1|\sigma _1,\sigma _2)w_2(\sigma _1,\sigma _2|\sigma _6,\sigma _7,\sigma
_8,\sigma _9)\times 
\]
\begin{equation}
\times \rho (\sigma _6,\sigma _7,\sigma _8,\sigma _9|\tau _2,\tau _3),
\label{23}
\end{equation}
where 
\[
w_2(\sigma _1,\sigma _2|\sigma _6,\sigma _7,\sigma _8,\sigma _9)= 
\]
\begin{equation}
=\sum_{\sigma _3}\sum_{\sigma _4}\sum_{\sigma _5}w(\sigma _1|\sigma
_3,\sigma _4)w(\sigma _2|\sigma _4,\sigma _5)w(\sigma _3|\sigma _6,\sigma
_7)w(\sigma _4|\sigma _7,\sigma _8)w(\sigma _5|\sigma _8,\sigma _9),
\label{24}
\end{equation}
and 
\begin{equation}
\rho (\sigma _6,\sigma _7,\sigma _8,\sigma _9|\tau _2,\tau _3)=\frac 1{%
\widetilde{P}(\tau _2,\tau _3)}R(\tau _2|\sigma _6,\sigma _7)R(\tau
_3|\sigma _8,\sigma _9)P(\sigma _6,\sigma _7,\sigma _8,\sigma _9),
\label{25}
\end{equation}
with 
\begin{equation}
\widetilde{P}(\tau _2,\tau _3)=\sum_{\sigma _6}\sum_{\sigma _7}\sum_{\sigma
_8}\sum_{\sigma _9}R(\tau _2|\sigma _6,\sigma _7)R(\tau _3|\sigma _8,\sigma
_9)P(\sigma _6,\sigma _7,\sigma _8,\sigma _9).  \label{26}
\end{equation}

To actually use the above equations we have to calculate in each step of the
renormalization the stationary probability distribution $P(\sigma )$ related
to the old transition probability $w(\sigma _i|\sigma _i^{\prime },\sigma
_{i+1}^{\prime })$ which satisfies the balance equation 
\begin{equation}
P(\sigma )=\sum_{\sigma ^{\prime }}W(\sigma |\sigma ^{\prime })P(\sigma
^{\prime }),  \label{31}
\end{equation}
where $W(\sigma |\sigma ^{\prime })$ is given by (\ref{1}). Of course, this
equation cannot be solved exactly and we should seek for approximate
solutions.

The simplest approximate solution of (\ref{31}) is obtained by writing the
following equation for the one-site probability 
\begin{equation}
P(\sigma _1)=\sum_{\sigma _1^{\prime }}\sum_{\sigma _2^{\prime }}w(\sigma
_1|\sigma _1^{\prime },\sigma _2^{\prime })P(\sigma _1^{\prime },\sigma
_2^{\prime }),  \label{32}
\end{equation}
and by inserting, in the right hand side of it, the following approximation
for the two-site probability 
\begin{equation}
P(\sigma _1,\sigma _2)=P(\sigma _1)P(\sigma _2).  \label{33}
\end{equation}
The equation becomes then a closed equation which can then be solved by
repeated iterations.

Better approximations can be set up by generalizing this procedure. An
approximation of order $n$ is obtained as follows. From equation (\ref{31})
we write down an equation for the $n$-site probability distribution, namely 
\[
P(\sigma _1,\sigma _2,...,\sigma _n)= 
\]
\begin{equation}
=\sum_{\sigma _1^{\prime }}...\sum_{\sigma _n^{\prime }}\sum_{\sigma
_{n+1}^{\prime }}w(\sigma _1|\sigma _1^{\prime },\sigma _2^{\prime
})w(\sigma _2|\sigma _2^{\prime },\sigma _3^{\prime })...w(\sigma _n|\sigma
_n^{\prime },\sigma _{n+1}^{\prime })P(\sigma _1^{\prime },\sigma _2^{\prime
},...\sigma _n^{\prime },\sigma _{n+1}^{\prime }).  \label{34}
\end{equation}
Next we use the following approximation \cite{dictom} \cite{avra} for the ($%
n+1$)-site probability 
\begin{equation}
P(\sigma _1,\sigma _2,\sigma _3,...,\sigma _n,\sigma _{n+1})=\frac{P(\sigma
_1,\sigma _2,...,\sigma _n)P(\sigma _2,\sigma _3,...,\sigma _{n+1})}{%
P(\sigma _2,\sigma _3,...,\sigma _n)},  \label{36}
\end{equation}
and insert it in the right hand side of (\ref{34}). Then equation (\ref{34})
becomes a closed equation for the $n$-site probability distribution since 
\begin{equation}
P(\sigma _2,\sigma _3,...,\sigma _n)=\sum_{\sigma _1}P(\sigma _1,\sigma
_2,...,\sigma _n).  \label{37}
\end{equation}

If the order of approximation $n\geq 4,$ then $P(\sigma _1,\sigma _2,\sigma
_3,\sigma _4)$ is obtained from $P(\sigma _2,\sigma _3,...,\sigma _n)$ by
summing over all $\sigma _i$ except $\sigma _1,$ $\sigma _2,$ $\sigma _3,$
and $\sigma _4.$ When $n<4,$ then for $n=1,$ 
\begin{equation}
P(\sigma _1,\sigma _2,\sigma _3,\sigma _4)=P(\sigma _1)P(\sigma _2)P(\sigma
_3)P(\sigma _4),  \label{38}
\end{equation}
for $n=2,$ 
\begin{equation}
P(\sigma _1,\sigma _2,\sigma _3,\sigma _4)=\frac{P(\sigma _1,\sigma
_2)P(\sigma _2,\sigma _3)P(\sigma _3,\sigma _4)}{P(\sigma _2)P(\sigma _3)},
\label{39}
\end{equation}
and for $n=3,$ 
\begin{equation}
P(\sigma _1,\sigma _2,\sigma _3,\sigma _4)=\frac{P(\sigma _1,\sigma
_2,\sigma _3)P(\sigma _2,\sigma _3,\sigma _4)}{P(\sigma _2,\sigma _3)}.
\label{40}
\end{equation}

\section{Results}

The renormalization transformation defined in the previous sections may be
sought of as a trajectory in the space spanned by the parameter $p_1,$ and $%
p_2.$ It can be easily checked that $\widetilde{w}(1|0,0)=0$ implying that a
system with one absorbing state renormalizes into another one with the same
property. Within the space of parameters $p_1$ and $p_2,$ the
renormalization transformation may be seen as a mapping $(p_1,p_2)%
\rightarrow (p_1^{\prime },p_2^{\prime }).$ Given $w(1|10)=w(1|01)=p_1,$ $%
w(1|11)=p_2,$ and $w(1|0,0)=0$ we obtain $p_1^{\prime }=\widetilde{w}(1|10)$
and $p_2^{\prime }=\widetilde{w}(1|11)$ by using equation (\ref{23}). It can
be easily checked that $\widetilde{w}(1|0,0)=0$ implying that a system with
one absorbing state indeed renormalizes into another one with the same
property.

We have iterated equation (\ref{23}) numerically using approximations of
order $n=1,$ $2,$ $3,$ $4,$ $5$ and $8.$ For all orders of approximation
that we used we have found, in the plane $(p_1,p_2),$ two fully attractive
fixed points $(0,0)$ and $(1,1),$ one fully repulsive fixed point $(1/2,1)$
and one nontrivial fixed point $(p_1^{*},p_2^{*})$. The attractive fixed
point $(0,0)$ is related to the absorbing state whereas the other attractive
fixed point $(1,1)$ is related to the active state. Almost all trajectories
are attracted to either one or the other of these two points as can be seen
in figure 1. The base of attraction are separated by a line, which should be
identified as the critical line of the Domany-Kinzel model. The separatrix
hits the line $p_2=1$ at $p_1=1/2,$ which is a fully repulsive fixed point,
and the line $p_2=0$ at $p_1=p_c.$

The nontrivial fixed point is located over the separatrix line and is
attractive along a direction parallel to the separatrix and repulsive in the
direction perpendicular (a hyperbolic fixed point). According to the RG
theory the critical exponent $\nu _{\perp },$ related to the correlation
length, is obtained from the eigenvalue $\lambda $ along the repulsive
direction through $\nu _{\perp }=\ln b/\ln \lambda .$ The eigenvalue $%
\lambda $ is the largest eigenvalue of the Hessian matrix related to the
transformation $(p_1,p_2)\rightarrow (p_1^{\prime },p_2^{\prime })$
calculated at the nontrivial fixed point. The location of the critical line
(the separatrix) is in fair agreement with previous results coming from
numerical simulation \cite{zebende}, as can be seen in figure 2. The
critical value $p_c=0.792$ of $p_1$ at $p_2=0$ differs from the value $%
p_c=0.799\pm 0.002$ obtained from numerical simulations \cite{zebende} in
about 1\%.

Table 1 shows the values of the critical point $p_c$ (along $p_2=0$) for
several values of the order of approximation up to $n=8.$ For $n\geq 4$ the
figures are very close to the value obtained from simulations. In the same
table it is shown the values of the nontrivial fixed point together with the
eigenvalue $\lambda $ and the critical exponent $\nu _{\perp }=\ln 2/\ln
\lambda .$ All figures for the eigenvalue $\lambda $ were obtained
numerically since it is very difficult to calculated the derivatives of the
Hessian matrix related to the transformation $(p_1,p_2)\rightarrow
(p_1^{\prime },p_2^{\prime })$. The best value of the critical exponents,
that were obtained from the approximation of order $n=8,$ is $\nu _{\perp
}=1.013$ which should be compared to $\nu _{\perp }=1.100\pm 0.005$ obtained
from transfer matrix methods \cite{dom}.

\section{Conclusion}

We used a real space RG to obtain the critical behavior of the Domany-Kinzel
cellular automaton. The phase diagram is the same as the one obtained from
simulation. The critical line, with the exception of the point $(1/2,1)$, is
dominated by just one nontrivial fixed point implying, within the ideas of
RG, that all critical points along the line are in the same universality
class, as expected. The location of the critical line is in good agreement
with previous results, the value of $p_1$ at $p_2=0$ differing in about 1\%
from the value obtained from numerical simulations. The best value of the
critical exponents is in fair agreement with the values obtained from
transfer matrix methods and from numerical simulations. A way of improving
these figures is to use a renormalization scheme for which the
renormalization factor $b$ is less than two. This could be accomplished, for
instance, by renormalizing a block of three sites into a block of two in
which case we would have $b=3/2.$ Another way of getting better values is to
use a RG space with more parameters instead of just two as we have done here.

\newpage\

\newpage\ 

\begin{center}
\textbf{Table caption}
\end{center}

\bigskip\ 

Table 1. For each value $n$ of the order of approximation it is shown the
critical point $p_c$ (along $p_2=0$), the nontrivial fixed point $%
(p_1^{*},p_2^{*}),$ the eigenvalue $\lambda $ corresponding to this fixed
point along the relevant direction and the critical exponent $\nu _{\perp
}=\ln 2/\ln \lambda ,$ related to the correlation length.

\newpage\ 

\begin{center}
\textbf{Table 1}
\end{center}

\smallskip\ 
\[
\begin{array}{llllll}
n & p_c & p_1^{*} & p_2^{*} & \lambda & \nu _{\perp } \\ 
1 & 0.745782 & 0.595189 & 0.826419 & 2.118 & 0.924 \\ 
2 & 0.778384 & 0.598983 & 0.927874 & 2.144 & 0.909 \\ 
3 & 0.788641 & 0.591302 & 0.949335 & 2.034 & 0.976 \\ 
4 & 0.792203 & 0.591732 & 0.946004 & 2.000 & 1.000 \\ 
5 & 0.791504 & 0.591872 & 0.945275 & 1.988 & 1.009 \\ 
8 & 0.791654 & 0.591930 & 0.944997 & 1.982 & 1.013
\end{array}
\]

\newpage\ 

\begin{center}
\textbf{Figure Caption}
\end{center}

\bigskip\ 

Figure 1. Renormalization group trajectories in the parameter space $%
(p_1,p_2)$ of the Domany-Kinzel cellular automaton, obtained by using the
approximation of order $n=4.$ Almost all trajectories flow either to one of
two fully atractive fixed points. One of them is related to the active state
(\emph{open circle}) and the other to the absorbing state (origin of
coordinates). The two base of atraction are separated by a line, identified
with the critical line of the model, where the hyperbolic nontrivial fixed
point (\emph{full circle}) and a fully repulsive point (\emph{star}) are
located.

\medskip\ 

Figure 2. Phase diagram of the Domany-Kinzel probabilistic cellular
automaton showing the frozen absorbing state and the active state. 
The squares are results coming from numerical simulations \cite
{zebende} and the line is the result obtained from the present real space
renormalization group using the approximation of order $n=4.$

\end{document}